\documentclass[11pt]{article}

\usepackage[utf8]{inputenc}
\usepackage{graphicx}
\usepackage{amsmath}
\usepackage{hyperref}
\hypersetup{
    colorlinks=true,
    linkcolor=blue,
    filecolor=magenta,      
    urlcolor=blue,
}

\usepackage{amsfonts}
\usepackage{setspace}
\usepackage{bbold}
\usepackage{appendix}
\usepackage{xcolor}
\usepackage{geometry}
\usepackage{tabularx}
\usepackage{caption}
\usepackage{subcaption}
\usepackage{multirow}
\newcommand{\cmmnt}[1]{\ignorespaces}

\usepackage[compact]{titlesec}
\titlespacing{\section}{0pt}{1ex}{1ex}
\titlespacing{\subsection}{0pt}{1ex}{0ex}
    
\title{Graph Signal Processing and Deep Learning: \\ Convolution, Pooling, and Topology}
\author{Mark Cheung, John Shi, Oren Wright, Lavender Y. Jiang, Xujin Liu, Jos\'e M.F. Moura\thanks{This material is based upon work partially funded and supported by the Department of Defense under Contract No. FA8702-15-D-0002 with Carnegie Mellon University for the operation of the Software Engineering Institute, a federally funded research and development center; DM20-0590. This work is also partially supported by NSF grants CPS~1837607 and CIF~1513936.}}
\date{July 29, 2020}

\usepackage{etoolbox}  
\AtBeginEnvironment{thebibliography}{\scriptsize}

\usepackage{lipsum}

\let\OLDthebibliography\thebibliography
\renewcommand\thebibliography[1]{
  \OLDthebibliography{#1}
  \setlength{\parskip}{0pt}
  \setlength{\itemsep}{0pt plus 0.3ex}
}

\begin{document}
\abovedisplayskip=5pt
\abovedisplayshortskip=5pt
\belowdisplayskip=5pt
\belowdisplayshortskip=5pt

\maketitle

\begin{abstract}
Deep learning, particularly convolutional neural networks (CNNs), have yielded rapid, significant improvements in computer vision and related domains. But conventional deep learning architectures perform poorly when data have an underlying graph structure, as in social, biological, and many other domains.  This paper explores 1)~how \textit{graph signal processing}~(GSP) can be used to extend CNN components to graphs in order to improve model performance; and 2)~how to design the graph CNN architecture based on the \textit{topology} or \textit{structure} of the data graph.

\end{abstract}

\section{Introduction}
\label{sec:intro}
\begin{figure}[htbp]
    \centering
	\includegraphics[height=6cm]{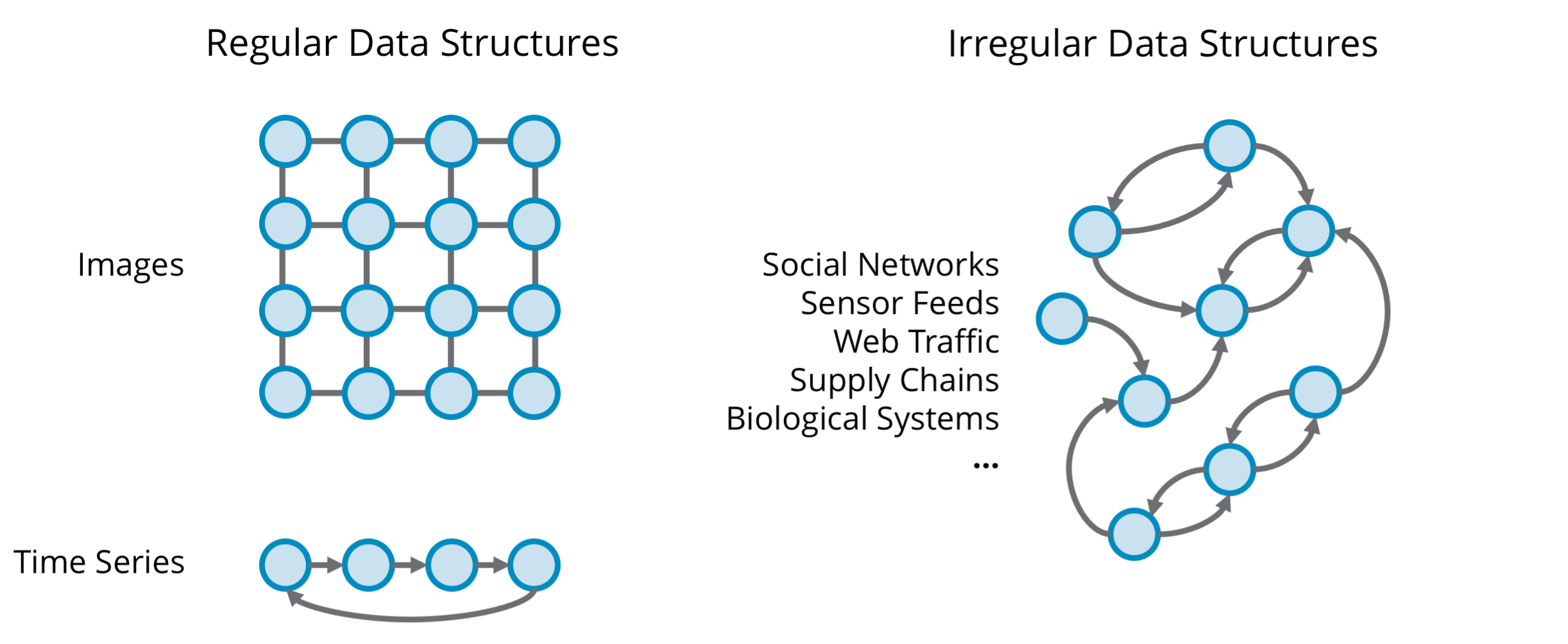}
	\caption{Problems in Euclidean data domains, like images and time series (left), have seen great advances through deep learning. Non-Euclidean data domains (right) require new deep learning techniques to achieve similar advances.}
	\label{fig:data_structures}
\end{figure}

Deep learning techniques such as convolutional neural networks (CNNs) have had a major impact on fields like computer vision and other Euclidean data domains (in which data have a uniform, grid-like structure), yet in many important applications data are indexed by irregular and non-Euclidean structures, which require graphs or manifolds to explicitly model. Such applications include social networks, sensor feeds, web traffic, supply chains, and biological systems. An illustration of these contrasting data structures is shown in Fig.~\ref{fig:data_structures}. 

Conventional deep learning approaches are often limited when the data index lacks Euclidean structure to exploit. A CNN that takes advantage of the indexing graph structure significantly outperforms CNNs that do not meaningfully incorporate that structure but are otherwise the same. The extension of deep learning to non-Euclidean data is an area of research now called \textit{geometric deep learning} \cite{bronstein_geometric_2017}. Our paper concentrates on 1)~how graph signal processing~(GSP) can be used to adapt the CNN architecture to take advantage of the graph structure, and 2)~how to design the resulting graph CNN architecture based on the graph structure.

Graph signal processing (GSP), the generalization of traditional digital signal processing to graphs \cite{DSP, ShumanNFOV, ortega_graph_2017}, can be used to extend CNNs to graph data. We concentrate on CNNs in this paper in part because of their particular effectiveness among deep learning models---notably because of the shift invariance of convolutional filters---and in part because the convolutional kernel is straightforward to define in GSP.

Graph structure, defined by its adjacency matrix, may differ from problem domain to problem domain to a far greater extent then, e.g., image structure in different computer vision problems. We consider the effect of graph structure with respect to 1)~\textit{node classification}, a semisupervised learning task in which unlabeled nodes of a graph are classified based on labeled nodes in the same graph, and 2)~\textit{graph classification}, a supervised learning task in which previously unseen graphs are classified based on a collection of labeled graphs. Analysis of the underlying graph structure can help in understanding how certain graphs may be easier to infer over, or are more suitable to different learning architectures.

\textbf{Remark:} The scope of this paper is on graph CNNs and two related applications: node and graph classification. There are many other geometric deep learning architectures not treated herein, such as graph autoencoders, recurrent graph neural networks, and spatial-temporal graph neural networks (e.g., \cite{li2019forecaster}). Please see \cite{review1, review2} for comprehensive surveys of graph neural networks. Also, there are important geometric deep learning problems beyond node classification and graph classification, including representation learning, link prediction, anomaly detection, graph generation, community detection, graph embedding, and combinatorial optimization (which require more complex architectures than graph CNNs). We consider only a small subset of application areas, mostly in chemistry and social networks. Some others include natural language processing (text generation, machine translation), computer vision (image classification, object detection), logistics, neuroscience, and other areas of science (see \cite{review1,review2}).

\textbf{Summary}: Below, we give a cursory introduction to CNNs and GSP in Section~\ref{sec:CNN} and Section~\ref{sec:GSP}, respectively. Section~\ref{sec:GCNN} addresses how GSP can be used with CNNs, leading to \textit{graph} CNNs, by discussing each of the core components of a graph CNN architecture. In Section~\ref{sec:Use}, we explore design considerations for graph CNNs; how the topology of the data graph defined by the graph adjacency matrix affects the performance of these models for node and graph classification. We then evaluate the results in Section~\ref{sec:res} and conclude with a discussion of open problems in Section~\ref{sec:Conc}.

\section{A Brief Overview of CNNs}
\label{sec:CNN}

\begin{figure}[ht]
    \centering
    \includegraphics[width=0.8\linewidth]{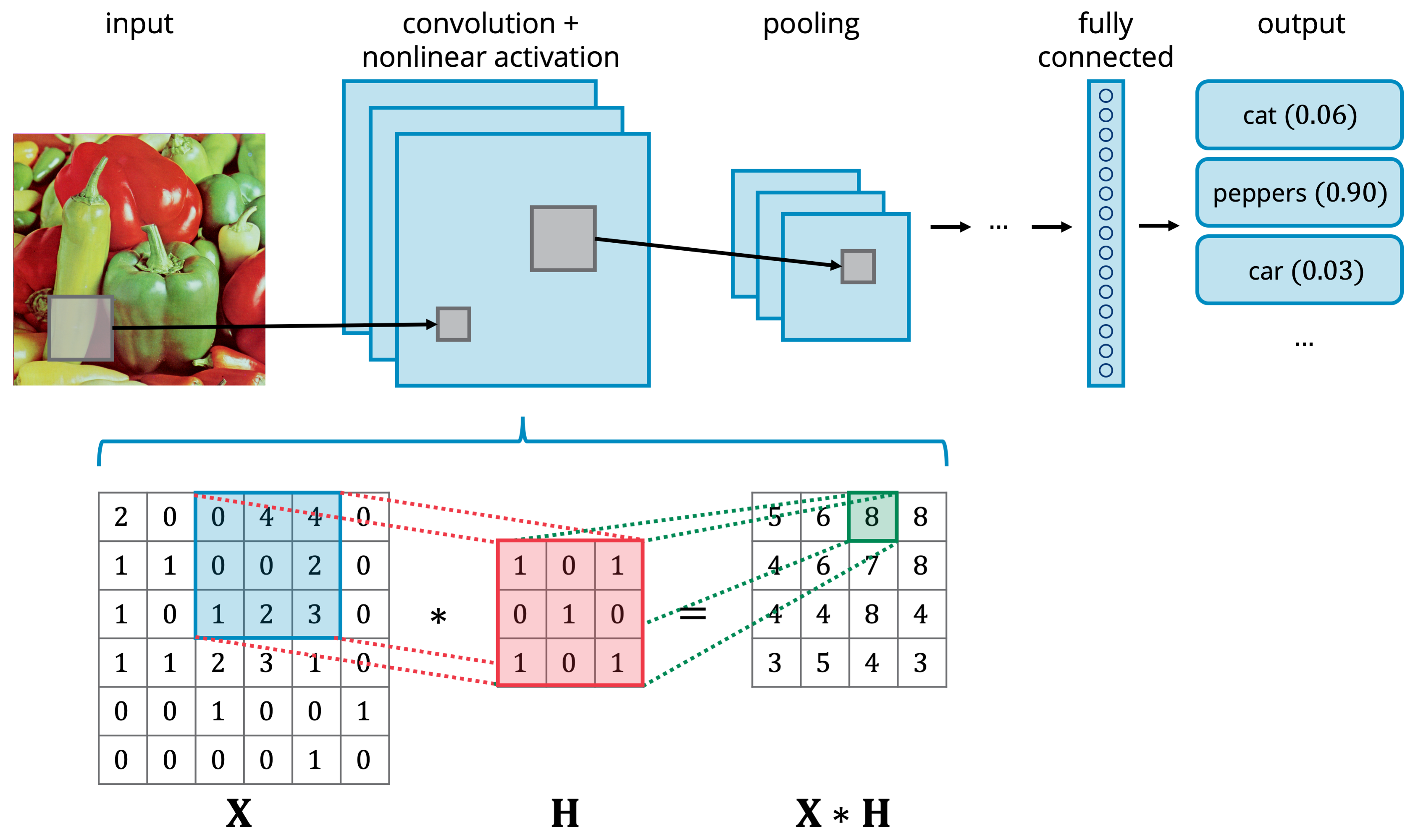}
    \caption[CNN Architecture and kernel]{Basic CNN architecture and kernel. A typical CNN consists of several component types (e.g., several convolutional + nonlinear activation layers, interleaved with pooling layers, followed by fully connected layers before prediction). The convolutional kernel sums the point-wise multiplications of a subset of the signal\footnotemark ~$\mathbf{X}$ with a learned filter $\mathbf{H}$. This operation is repeated as the filter ``slides'' across the input. (Note that what is termed convolution by the deep learning community is actually cross-correlation.)}
    \label{fig:cnn}
\end{figure}

CNNs, like that shown in Fig.~\ref{fig:cnn}, have proven effective in domains like computer vision because the convolutional kernel has several powerful properties: 1) it has a fixed number of parameters, allowing for a small memory footprint and computational cost; 2) it operates locally, meaning higher-level, global features can be composed of lower-level, local features; and 3) it is shift invariant. In constructing a graph CNN, such properties should ideally translate. The basic operation of a CNN convolutional kernel is shown in detail in Fig.~\ref{fig:cnn}, where a $3 \times 3$ filter or kernel (red window) slides over the input ($\mathbf{X}$) to generate the output.

The CNN architecture has been successfully applied to many classification tasks like image classification and speech recognition. The model shown in Fig.~\ref{fig:cnn} transforms an input through a sequence of hidden layers. Each early-stage hidden layer consists of 1) a set of learned filters like convolution, followed by 2) a nonlinear activation function, and 3) pooling. A filter outputs a feature map that serves as the input to the subsequent layer, which also includes several parallel channels (3 shown in Fig.~\ref{fig:cnn}). Later hidden layers are typically fully connected and, finally, a prediction is made at the output layer, using a loss function like cross-entropy loss.  Numerous variants and improvements have been detailed across the literature. A more complete treatment of CNNs can be found in \cite{LeCun2015}.

\footnotetext{For CNNs, we denote the signal as $\mathbf{X}$ because the input is typically an image and therefore a matrix.}

\section{Elements of Graph Signal Processing}
\label{sec:GSP}
 Graph signal processing (GSP) extends traditional signal processing operations such as shifting, Fourier transforms, convolution, and sampling to graphs \cite{DSP, ShumanNFOV, ortega_graph_2017}. GSP provides an intuitive, theoretically rigorous framework to evaluate signals on graphs and can be used to extend CNNs to graphs. In this section, we give a brief primer on GSP theory, which is related to the graph CNN architecture and analysis in the subsequent sections. More thorough introductions to GSP can be found in \cite{DSP, ShumanNFOV, ortega_graph_2017}.

\subsection{The Shift Operator}
\label{sec:shift}

GSP has been developed to process, from first principles, graph-structured data \cite{DSP, ShumanNFOV, ortega_graph_2017}. GSP can be thought of as a generalization of classical signal processing: whereas the structure indexing classical signals is implicit, the structure of non-Euclidean data must be explicitly represented, which GSP does with the pairwise relationships of graphs. Because GSP is a generalization of classical signal processing \cite{DSP, ShumanNFOV, ortega_graph_2017}, GSP reduces to DSP in the special case of a uniform graph (e.g., the pixels of an image \cite{puschel2008} or indices of a time series).

GSP can perhaps best be understood by beginning with a generalization of the shift operation from DSP. A finite-support (or periodic) discrete-time signal $x[n]$ with period $N$ can be represented by a vector $x=[x_0, x_1, \dots, x_{N-2}, x_{N-1}]^T$. In this representation, to filter $x[n]$ by some finite impulse response (FIR) filter $g$, we represent $g$ as a matrix $\mathbf{G}$ and simply perform a matrix multiplication $x' = \mathbf{G}x$.

The shift operator plays a crucial role in DSP; any linear, shift-invariant filter, \textbf{G}, can be expressed as a polynomial of the circular shift. For the time model and DSP, we can write the circular shift as
\begin{equation}
\label{eq:A}
    \mathbf{A} =
\begin{bmatrix}
	0  & 0 & 0 & \hdots & 0 & 1 \\
	1  & 0 & 0 &  \hdots & 0 & 0 \\
	0  & 1 & 0 &  \hdots & 0 & 0 \\
	\vdots  & \vdots & \ddots & \ddots & \vdots  & \vdots\\
	0  & 0 & 0 &  \hdots & 1 & 0 \\
	\end{bmatrix}.
\end{equation}
Given this choice for the shift operator~$\mathbf{A}$, the shift operation\footnote{Shifting the signal $x[n]$ to the right to produce $x[n-1]$.} in DSP on $x$ is written as $\mathbf{A}x$, i.e.,
\begin{equation}
\label{eq:ring_shift}
    [x_{N-1}, x_0, \dots, x_{N-3}, x_{N-2}]^T = 
    \mathbf{A}[x_0, x_1, \dots, x_{N-2}, x_{N-1}]^T.
\end{equation}
We can interpret this DSP operation as a graph operation by recasting $\mathbf{A}$ as an adjacency matrix—a matrix representation of a graph. More precisely, $\mathbf{A}$ corresponds to the adjacency matrix of a directed ring graph $\mathcal{G}=(\mathcal{V,E})$, where $\mathcal{V}$ is a set of $N$ vertices and $\mathcal{E}$ the set of directed edges connecting each vertex to its next neighbor (with the final vertex connecting back to the first). The $N$ elements of the signal or data $x$ are indexed by the $N$ vertices of $\mathcal{V}$, and each element $(\mathbf{A})_{ij}$ is the weight $w_{ij}$ of the edge connecting vertex $i$ to $j$.

This dual role played by the matrix $\mathbf{A}$ is the key to constructing a linear, shift invariant GSP \cite{DSP}. Instead of using only the DSP ring graph, with adjacency matrix given by \eqref{eq:A}, we generalize this to an arbitrary graph in GSP.

We note here that graph shift operators other than $\mathbf{A}$ have been proposed, such as the graph Laplacian $\mathbf{L}=\mathbf{D}-\mathbf{A}$, where $\mathbf{D}$ is the degree matrix of $\mathcal{G}$, defined as the diagonal matrix $(\mathbf{D})_{ii}=\sum_j (\mathbf{A})_{ij}$. Graph Laplacians apply only to undirected graphs, since they are symmetric and positive semidefinite, and have been widely studied within the field of spectral graph theory \cite{ortega_graph_2017}.

\subsection{Graph Convolution}
\label{sec:graphconv}
Under appropriate conditions\footnote{The characteristic and minimal polynomials are the same. To simplify the discussion in the paper, we assume the sufficient condition that the eigenvalues of $\mathbf{A}$ are distinct.}, graph convolution is defined by the matrix vector multiplication
\begin{equation}
\label{eq:gconv}
    \mathbf{G}x = \text{g}(\textbf{A})x =  \sum_{k=0}^K \alpha_k \mathbf{A}^k x, \; K<N
\end{equation}
where $x$ is the graph data or graph signal, $\text{g}(\textbf{A})$ is the polynomial filter of degree $K$, and $\alpha_k$ the $k$-th filter coefficient. In practice, $\mathbf{A}$ is often normalized in some manner to ensure numerical stability. For example, dividing $\textbf{A}$ by $|\lambda_{\text{max}}|$ where $\lambda_{\text{max}}$ is the eigenvalue of $\textbf{A}$ with greatest magnitude, or, with an undirected graph, using $\mathbf{\bar{A}}=\mathbf{D}^{-\frac{1}{2}}\mathbf{A}\mathbf{D}^{-\frac{1}{2}}$ in place of $\mathbf{A}$, where $\mathbf{D}$ is the degree matrix of $\mathbf{A}$. These guarantee that the (non-maximal) eigenvalues of the adjacency matrix are inside the unit circle, thereby making $\mathbf{G}$ computationally stable.

If we reorder the nodes of $\textbf{A}$, then we get an adjacency matrix $\textbf{A}_{\textrm{new}} = \textbf{PAP}^T$, where $\textbf{P}$ is a permutation matrix. This also reorders the graph signal $x$ to become $x_{\textrm{new}}=\textbf{P}x$. Using \eqref{eq:gconv}, graph convolution with permuted graph signal and adjacency matrix becomes
\begin{equation}
    \label{eq:convperm}
    \mathbf{G}_{\textrm{new}}x_{\textrm{new}} = \text{g}(\mathbf{A}_{\textrm{new}})x_{\textrm{new}} = 
    \sum_{k=0}^K \alpha_k \mathbf{A}_{\textrm{new}}^k \textbf{P}x = 
    \sum_{k=0}^K \alpha_k \textbf{P}\mathbf{A}^k\textbf{P}^T \textbf{P}x =
    \textbf{P} \sum_{k=0}^K \alpha_k \textbf{A}^k x = \textbf{P}\left(\textbf{G}x\right).
\end{equation}
Thus, if we reorder the nodes and convolve, we obtain the original result reordered in the same manner. This is a desirable property for learning architectures because reordering the nodes in $\mathbf{A}$ should not adversely affect the inference result.

\subsection{Frequency Representation}

GSP unifies the vertex and spectral domains of a graph, much as classical signal processing connects the time and frequency domains of a time-series.

In DSP, the discrete Fourier transform (DFT)\footnote{$\text{DFT} = \frac{1}{\sqrt{N}} \left[e^{-\frac{2\pi jkl}{N}}\right]_{k,l = 0,1,\hdots,N-1}$, $j = \sqrt{-1}$. The DFT is unitary; i.e, $\text{DFT}^H = \text{DFT}^{-1}$.} follows from the eigendecomposition of the ring graph adjacency matrix  $\mathbf{A}$ given in \eqref{eq:A},
\begin{equation}
\label{eq:A_eigen}
    \mathbf{A} = \text{DFT}^H\mathbf{\Lambda}\ \text{DFT}.
\end{equation}
Similarly, in GSP, the graph Fourier transform (GFT) is defined by the eigendecomposition of an arbitrary adjacency matrix
\begin{equation}
\label{eq:A_eigengsp}
    \mathbf{A} = \text{GFT}^{-1}\mathbf{\Lambda} \ \text{GFT}.
\end{equation}
The inverse graph Fourier transform $\text{GFT}^{-1}$ is formed with the eigenvectors of $\textbf{A}$. $\text{GFT} = \text{DFT}$ for a directed ring graph where $\textbf{A}$ is in \eqref{eq:A}. The graph spectral representation of the graph signal $x$ is given by $\widehat{x} = \text{GFT}x$.

\section{Graph CNN Architecture}
\label{sec:GCNN}
\begin{figure}[ht]
    \centering
    \includegraphics[width=0.8\linewidth]{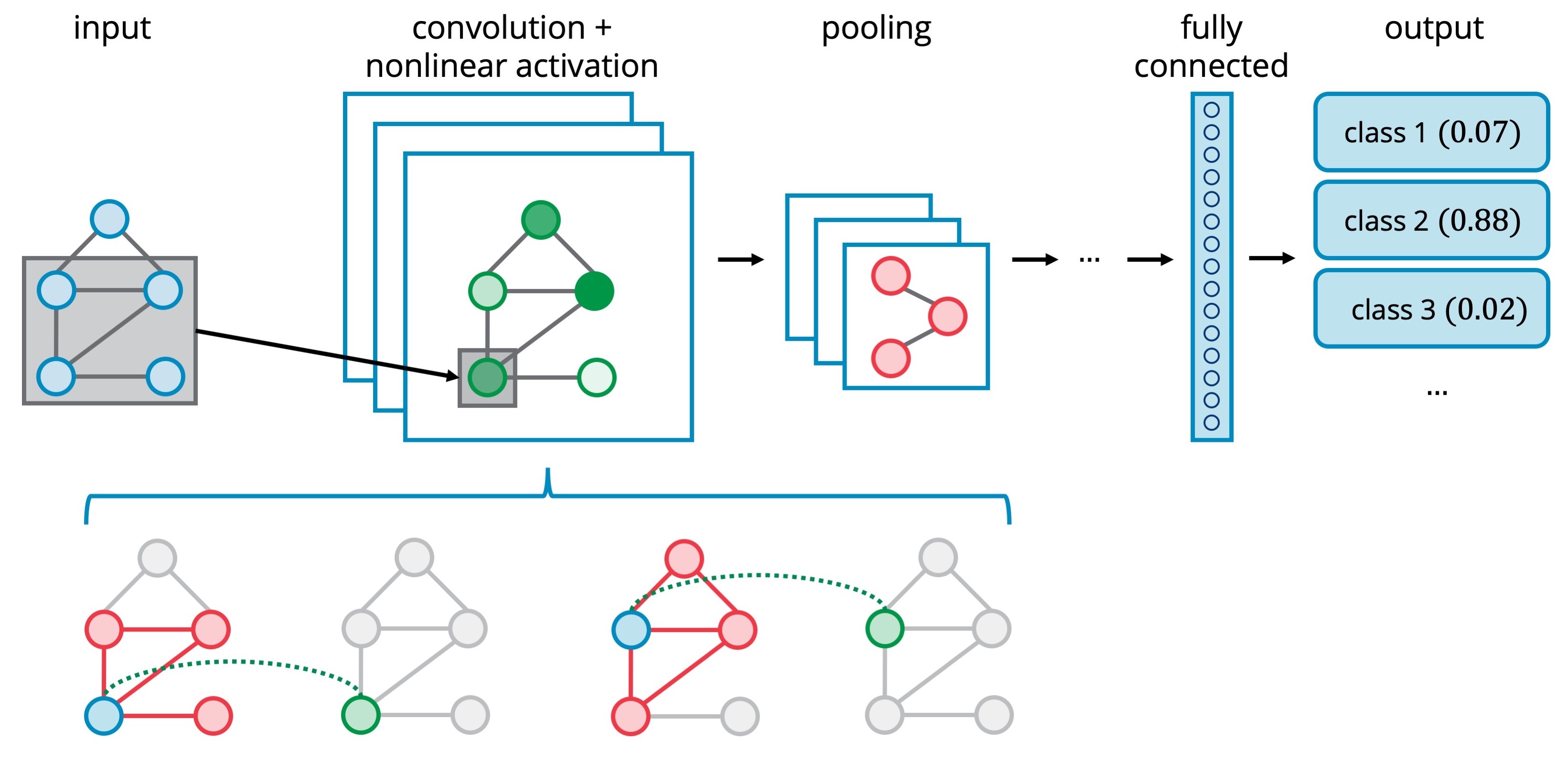}
    \caption{Graph CNN architecture and kernel, extending CNN elements to graph-structured data. An example graph convolutional filter of degree $1$ is shown. The filter, centered on the blue vertex, aggregates neighborhood information (shown in red) by multiplying the graph signal with a polynomial of shifts to produce an output (shown in green). The filter is applied on each node in the graph.}
    \label{fig:gcnn}
\end{figure}

Like a conventional CNN architecture, a graph CNN architecture also has convolutional, pooling, and fully connected layers (shown in Fig.~\ref{fig:gcnn}). Convolutional and pooling layers can be formulated anew to incorporate graph structure using GSP theory. We describe these layers in this section and explore design considerations in Section~\ref{sec:Use}.

\subsection{Graph Convolutional Layer}

\label{sec:graphCNN}

Broadly speaking, two approaches to graph CNNs have been pursued, the spectral and the vertex domain approaches. Graph convolution was first generalized from CNNs to graph-structured data in the spectral domain using the graph Laplacian. The spectral approach takes the graph signal $x$, multiplies by GFT to get $\widehat{x}$, and then multiplies by $\text{GFT}^{-1}$ to return to the vertex domain \cite{bronstein_geometric_2017, bruna_spectral_2013}. To avoid the expensive eigendecomposition of \textbf{A} to find the GFT matrix, \cite{defferrard_convolutional_2016} used Chebyshev polynomials to approximate the GFT.

The vertex approach defines convolution in the vertex domain, as given by \eqref{eq:gconv}. These approaches are typified graph convolutional networks (GCNs) \cite{kipf17} and topology-adaptive graph convolutional networks (TAGCNs) \cite{du2017}, which we consider below. Other implementations, like GraphSAGE \cite{hamilton2017} and graph attention network (GATs) \cite{Velickovic2018}, are deep learning approaches defined in the vertex domain, but their kernels do not meet the definition of convolution in GSP.

Consider a graph $\mathcal{G}=(\mathcal{V,E})$, where $\mathcal{V}$ is a set of $N$ vertices defined by graph signal\footnote{For ease of notation, we use $x$ here to refer to the graph signal. If each node has more than one channel ($C>1$), then $x$ can be a matrix.} $x^{(0)} \in \mathbb{R}^{N \times C}$($C$ is the number of signal dimensions, or channels) and $\mathcal{E}$ is defined by its adjacency matrix $\mathbf{A} \in \mathbb{R}^{N \times N}$. The GCN \cite{kipf17} convolutional layer in general form is
\begin{equation}
\label{eq:gcn}
    x^{(\ell+1)}=
    \sigma \left(
    \tilde{\mathbf{A}}
    x^{(\ell)} \mathbf{W}^{(\ell)} \right),
\end{equation}
where $\tilde{\mathbf{A}}=\mathbf{A}+\mathbf{I}_N$, $\mathbf{W}^{(\ell)} \in \mathbb{R}^{C \times F}$ is the trainable weight matrix, $\sigma$ is the nonlinear activation function, and $F$ is the number of output features. The layer number is $\ell$, with $\ell=0$ for the input layer. Adding the identity matrix to $\mathbf{A}$ and then normalizing helps address numerical instabilities and vanishing gradients.

The TAGCN implementation of graph convolution \cite{du2017} treats in addition to $\mathbf{W}^{(\ell)}$ the polynomial filter coefficients as learnable weights and their degrees as hyperparameters. We can write the general form of the TAGCN graph convolutional layer as
\begin{equation}
\label{eq:tagcn}
     x^{(\ell+1)} = \sigma \left( x^{(\ell)}\mathbf{W}_0^{(\ell)} + \textbf{A} x^{(\ell)} \mathbf{W}_1^{(\ell)}+ \hdots + \textbf{A}^K x^{(\ell)} \mathbf{W}_K^{(\ell)}
     \right) = \sigma \left(\sum_{k=0}^K \textbf{A}^k x^{(\ell)} \mathbf{W}_k^{(\ell)} \right),
\end{equation}

where $K$ is the degree of the graph polynomial filter.
A visual representation of the TAGCN convolutional layer is shown in Fig.~\ref{fig:gcnn}. A degree 1 graph filter (bottom) uses information from first-order neighbors to compute its output. The polynomial coefficients are learned, similar to the filter coefficients of a CNN. Like a CNN, the filter ``slides'' across the graph from vertex to vertex, and the output is fed to a nonlinear activation function. As mentioned in Section~\ref{sec:graphconv}, we can use normalized versions of $\tilde{\textbf{A}}$ and $\textbf{A}$ in \eqref{eq:gcn}, \eqref{eq:tagcn}.

The graph convolutional layer in GCN \cite{kipf17} and TAGCN \cite{du2017} can be stated in terms of the vertex domain covolution in \eqref{eq:gconv}. Each graph convolutional layer can be interpreted in GSP as: $x' = \sigma\left(\text{g}(\textbf{A}) x\right)$
where $x$ and $x'$ are the input and output of the layer, $\text{g}(\textbf{A})$ a polynomial of degree $K$ and $\sigma$ the nonlinear activation.
By~\eqref{eq:A_eigengsp}, the convolutional layer (ignoring nonlinear activation) can be interpreted in the spectral domain as $\text{g}(\mathbf{\Lambda})\widehat{x}$ with
\begin{equation}
    \label{eqn:Achain}
    \text{g}(\mathbf{A})x = \text{GFT}^{-1}\text{g}(\mathbf{\Lambda})\text{GFT}x = 
    \text{GFT}^{-1}\text{g}(\mathbf{\Lambda})\widehat{x} \xrightarrow{\text{GFT}}\text{g}(\mathbf{\Lambda})\widehat{x}.
\end{equation}

\subsection{Graph Pooling Layer}
\label{subsec:graphpoolinglayer-1}
Pooling in some form is usually desirable in graph classification models for two reasons: dimensionality reduction and hierarchical learning. Graph pooling algorithms generally reduce the number of nodes and hence the number of learned parameters of the model. Some pooling algorithms also enforce hierarchical representation of the data, so the graph CNN can learn large-scale and global patterns in the data.

\begin{figure}[htpb]
    \centering
    \includegraphics[width=0.8\linewidth]{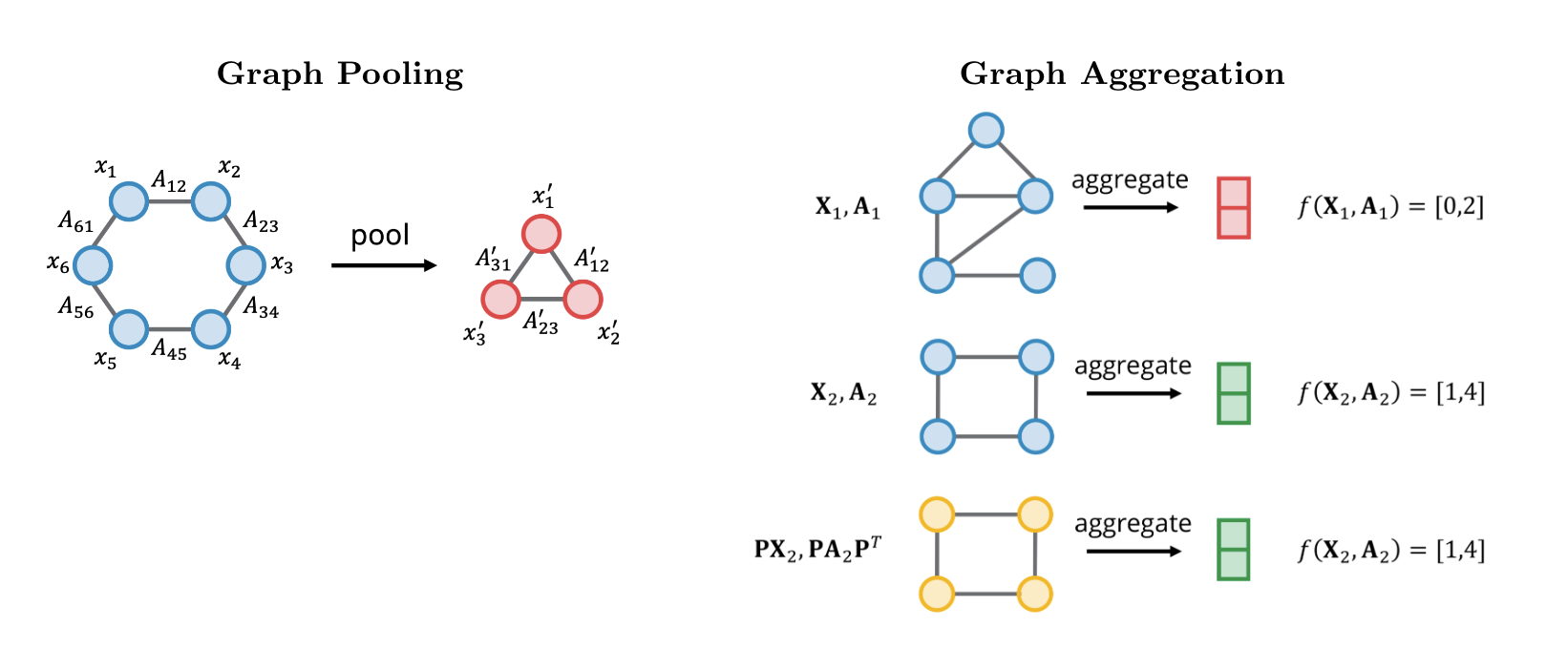}
    \caption{Graph pooling and graph aggregation. Graph pooling (left) accepts a graph signal and produces a new, representative graph signal indexed by a smaller graph support. Global aggregation (right) can accept graphs of potentially varying sizes and produce fixed-length, representative vectors.}
    \label{fig:pool_aggregation}
\end{figure}
\begin{figure}[hbtp]
    \centering
    \includegraphics[width=0.8\linewidth]{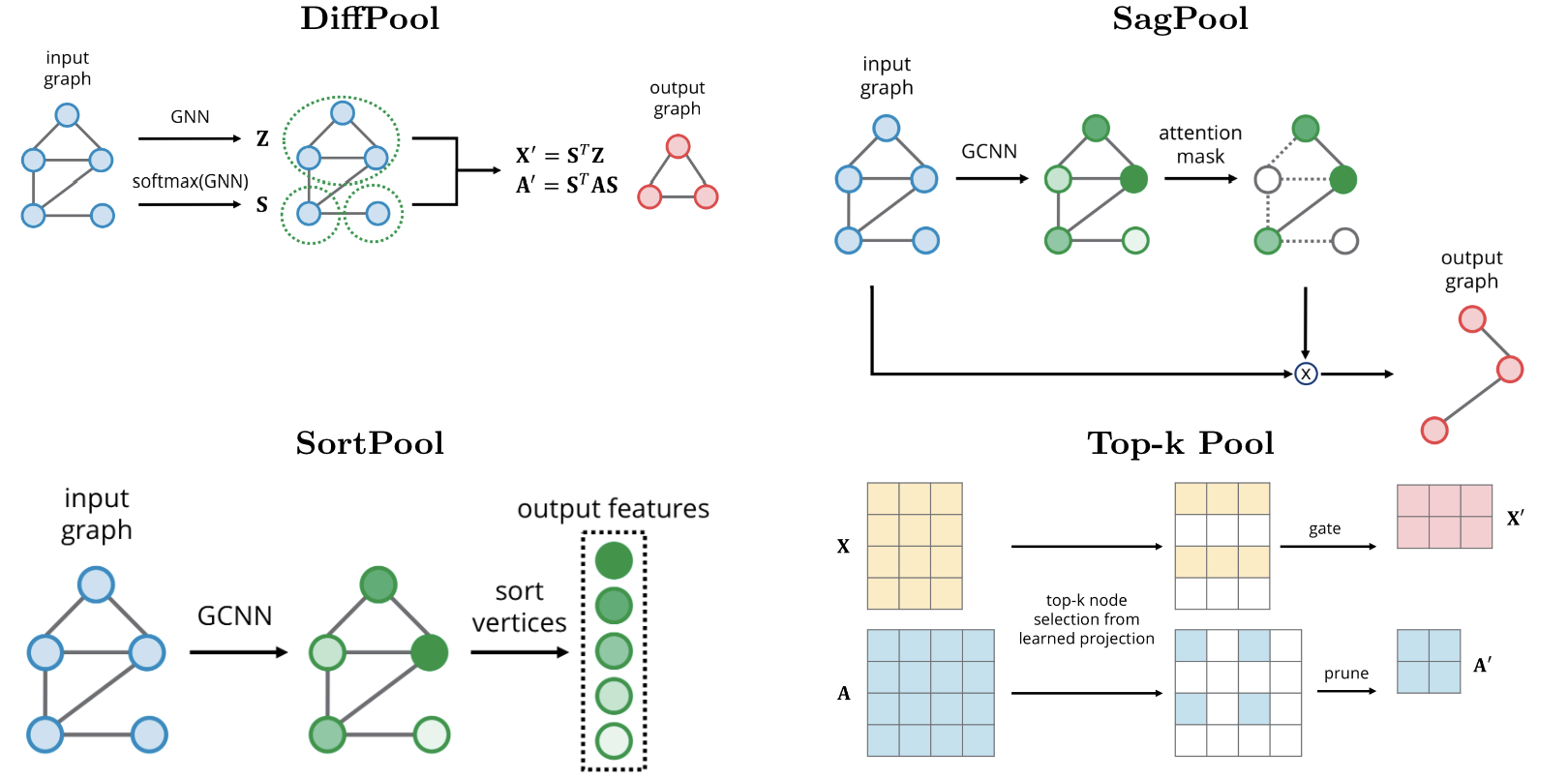}
    \caption{High-level illustrations of recently proposed graph pooling methods. DiffPool \cite{ying2018} uses a graph CNN model to obtain an assignment matrix for clustering the nodes. SagPool \cite{SagPool2019} uses a graph CNN layer to calculate self-attention as mask for pooling the nodes. In SortPool \cite{zhang2018}, graph CNN layers are followed by a ranking of the nodes to select the top nodes. Top-k Pool \cite{gao2019} uses a projection vector to select the the top nodes, which form a new graph.}
    \label{fig:graph_pooling_methods}
\end{figure}

More formally, using the notation in Section~\ref{sec:graphCNN}, graph pooling yields signal $x' \in \mathbb R^{N' \times C}$ and adjacency matrix $\mathbf{A}' \in \mathbb R^{N' \times N'}$, with $N' \leq N$ (shown in Fig.~\ref{fig:pool_aggregation}). Pooling can be understood as a nonlinear downsampling operation. In deep learning, unlike as normally treated in DSP, recoverability is not a key concern for downsampling, and unlike graph convolution, which has a GSP definition, graph pooling has not been rigorously defined. In CNNs, max pooling is usually used, whereas in graph CNNs, there is no consensus how best to pool nodes in a graph. Recent methods of graph pooling include Sort Pooling (SortPool) \cite{zhang2018}, Differentiable Pooling (DiffPool) \cite{ying2018}, Top-k Pool \cite{gao2019}, and Self-Attention Graph Pooling (SagPool) \cite{SagPool2019}. See Fig.~\ref{fig:graph_pooling_methods} for an overview. In Section~\ref{sec:graphsample}, we discuss pooling's connection to sampling in GSP as an open problem.

\subsection{Graph Aggregation Layer}
\label{subsec:graphaggregationlayer}
In conventional CNNs, inputs are generally of the same size and have a fixed ordering. However, in graph classification problems, one often needs to consider graphs of various sizes, and with data defined on each graph in an arbitrary node-permuted order (e.g., molecular networks). The resulting vectors that would serve as input to a fully connected layer vary in both size and labeling order, making direct comparison difficult.

The graph aggregation layer, sometimes called the final pooling layer, solves this problem by collapsing the nodes into a fixed number of features, regardless of input size, for comparison (see Fig.~\ref{fig:pool_aggregation}). This is often done using a mean or sum operation over all nodes in a graph. However, several other statistics and approaches can be used. One such heuristic is the family of graph spectral distances, $\textrm{F}_{\textrm{GSD}}$ \cite{verma2017}, which uses the adjacency matrix to capture global information about the graph structure by taking harmonic distances for all nodes. Other approaches include using graph capsules (e.g., GCAPs-CNN \cite{verma2018} and CapsGNN \cite{xinyi2019})

\section{Design Considerations and Applications}
\label{sec:Use}
In this section, we address how the structure of the graph $\mathcal{G}$ defined by its adjacency matrix $\mathbf{A}$ affects the design of graph CNNs for two application areas: node and graph classification. The key difference between graph CNNs and traditional CNNs is how the kernel operations in each layer are implemented. Conventional deep learning algorithms can still be applied (e.g., backpropagation, dropout, and gradient-based optimization). We consider the design of three components of graph CNN architecture: graph convolutional layer, graph pooling layer, and graph aggregation layer.

\subsection{Node Classification}
\label{sec:node_classification}
\begin{figure}[htbp]
    \centering
    \includegraphics[width=\linewidth]{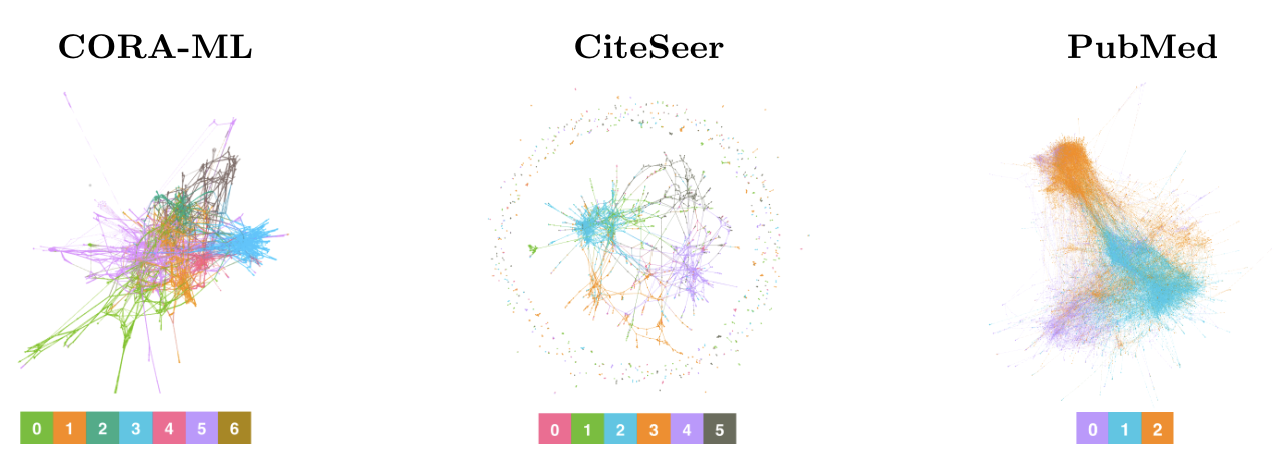}
    \caption{Visualization \cite{GEPHI} of citation network datasets. Class labels are color-coded.}
    \label{fig:viz}
\end{figure}

In node classification, we infer the classes of unlabeled nodes in a graph using the graph's labeled nodes. Common node classification benchmarks are the citation networks, three of which \cite{node_dataset} are visualized in Fig.~\ref{fig:viz}: CORA-ML, CiteSeer and PubMed. On these graphs, each node represents a scientific publication, with bag-of-words feature vectors. An undirected edge is formed between two nodes if one cites the other. Node labels represent different publication categories. In CORA-ML, for example, the labels represent seven subfields of machine learning (e.g., computational biology and natural language processing).

\begin{figure}[htbp]
    \centering
    \includegraphics[height=5cm]{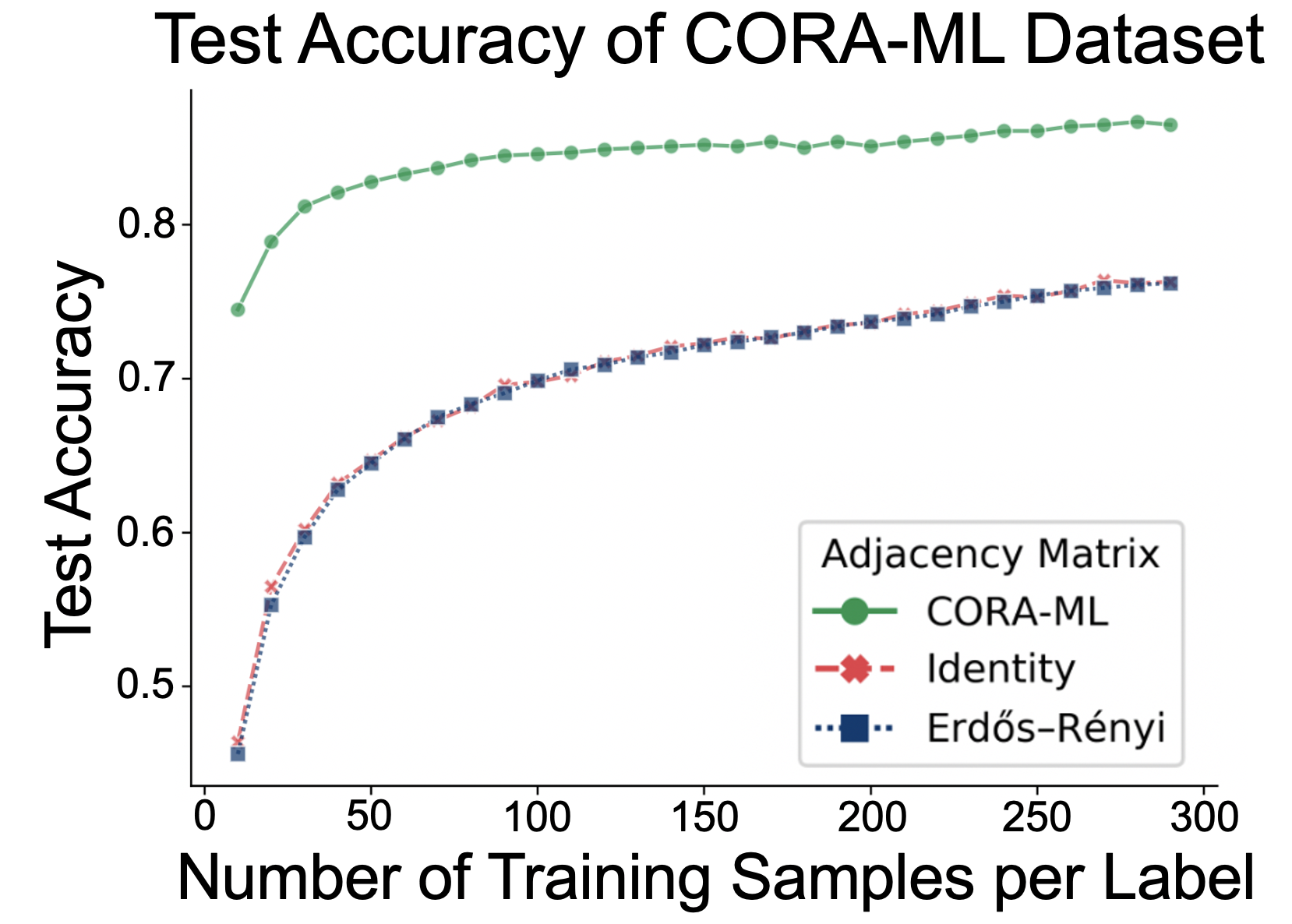}
    \caption{Comparison of graph CNN test accuracy on the CORA-ML citation network dataset with three types of graph structures: 1)~the dataset's citation network graph, CORA-ML, 2)~the identity matrix, and 3)~an Erd\H{o}s-R\`enyi graph.}
    \label{fig:graphEffect}
\end{figure}

\textbf{The effect of representing graph structures:} We first confirm that accounting for the graph structure defined by the adjacency matrix in the neural network significantly improves performance. In Fig.~\ref{fig:graphEffect}, we consider the performance of GCN on the CORA-ML dataset (its graph structure is visualized in Fig.~\ref{fig:viz}). The green trend shows GCN performance when trained with the graph structure of CORA-ML, which significantly outperforms the same model when the graph structure is ignored (i.e., trained with the identity matrix) or randomly generated (i.e., trained with a random Erd\H{o}s-Ren\`yi graph).

\textbf{Measuring graph structure effectiveness:} For a node classification task, it is natural to ask how useful the graph structure is. To measure this, we consider the graph structure, the number of classes, the label rate, and the content of the graph signal. To this end, we define a graph structure as \textit{effective} if a graph CNN model trained with the adjacency matrix  $\mathbf{A}$ has higher test accuracy than a graph CNN model trained with the identity matrix (which is effectively a neural network that does not incorporate the graph structure). To measure the effectiveness of graph structure a priori, we introduce a metric called edge entropy. Edge entropy measures the quality of the label information encoded in the graph based on the graph's conditional class distribution. A low edge entropy implies a more certain class distribution, and therefore a more useful graph structure for classification. Fig.~\ref{fig:viz} shows that CORA-ML has relatively low edge entropy, because nodes of the same class tend to cluster together. In such a case, the graph structure is helpful because the edges encode predictive label information. For example, if we know a node $u$ is a math paper, and node $v$ is a neighbor of node $u$, then it is likely that node $v$ is also a math paper.

Formally, given a graph $\mathcal{G}$ with~$M$ node classes, we define the $n$-th-order edge entropy of any class $i$ as a function $H_n: \{1, \hdots, M\} \to [0,1]$ such that 
\begin{equation}\label{edgeEntropy}
	H_{n}(i) := -\sum_{j \in \{1, \hdots, M\}} p_{ij}(n)\log_{M}(p_{ij}),
\end{equation}
where the $n$-th order interclass connectivity probability $p_{ij}$ is defined as
\begin{equation}\label{connProb}
p_{ij}(n) := \frac{|\{\text{length } n \text{ walk } w: \text{start}(w) \in \mathcal{V}_{i} \land \text{end}(w) \in \mathcal{V}_{j} \}|}{|\{\text{length } n \text{ walk } w: \text{start}(w) \in \mathcal{V}_{i}\}|},
\end{equation}
where $\mathcal{V}_{i}$ is the set of nodes that belong to the $i$-th class, $|\cdot|$ is the cardinality of the set, and $p_{ij}$ is the conditional class distribution of nodes that are a walk of length $n$ away from a node of class $i$ (i.e., $p_{ij}$ is the probability that a node~$v$ belongs to class~$j$ given that there exists a walk of length~$n$ between~$v$ and a node of class~$j$). More comprehensive results and analysis of edge entropy, including other parameters, are to be published elsewhere. 

\textbf{How to choose the graph CNN architecture:} Here, we consider the number of convolutional layers. A graph CNN architecture for node classification tasks has graph convolutional layers, but no pooling or aggregation layers (as each node need to be classified). We consider the two graph convolution variants, GCN \cite{kipf17} and TAGCN \cite{du2017}, defined formally in \eqref{eq:gcn} and \eqref{eq:tagcn}. Choosing the polynomial order of each graph convolutional layer is a design consideration for TAGCN, but not GCN. This is because a GCN convolutional layer has a fixed polynomial degree of 1 ($\mathbf{A} + \mathbf{I}$), whereas a TAGCN layer can have degree $K$. Choosing the number of graph convolutional layers is a design consideration for both variants. The objective of such considerations is to sufficiently capture the structure and complexity of the data without oversmoothing: where too much distinguishing information at each node is lost to repeated graph convolutions. Theoretically, if we choose the number of layers such that the overall order $K$ is the graph diameter (longest shortest path length) in the given graph structure, then the model can learn to combine the information of any two nodes. In practice, a small number of layers (e.g., 2) will perform better unless oversmoothing is addressed in some way \cite{oversmooth}. For citation networks, our results show that 2 layers generally give the best performance (see Section~\ref{subsec:graph_convolution_results}, Fig.~\ref{fig:node_classification_results}).

\subsection{Graph Classification}
\label{sec:graph_classification}

In graph classification, we infer the classes of unlabeled graphs, given a set of labeled graphs (with graph structures, graph signals, and class labels given a strict subset of graphs). This is analogous to the image classification task seen in computer vision. For example, in MUTAG \cite{graphstat}, molecules are graphs defined according to the chemical bonds between atoms and the task is to predict whether a molecule is mutagenic or not.

\textbf{How to choose the graph CNN architecture:} Here, we consider the graph convolutional layer, graph pooling layer, and aggregation layer. In graph classification, large-scale or global structure may be important. For example, predictive structural relationships may exist between large subgraph structures that are not visible between individual nodes or small neighborhoods. It is helpful to consider the sparsity or connectedness of the graphs in question using network parameters like average degree and graph diameter. These parameters can inform graph CNN architecture design.

We consider the two graph convolution variants: GCN \cite{kipf17} and TAGCN \cite{du2017}. As shown in \eqref{eq:gcn} and \eqref{eq:tagcn}, GCN considers the immediate neighbors of each node in each layer (one hop away), and  TAGCN, with a polynomial filter of degree $K$, considers nodes that are~$K$ hops away in each layer. In GCN convolution, after $\ell$ layers, each node has only received information from nodes~$\ell$ hops away. With polynomial filters of degree~$K$ in TAGCN, the output at the $i$-th node of the $\ell$-th layer depends on the input values of nodes up to $K\ell$ hops away. TAGCN presents an interesting tradeoff between number of layers and degree of the polynomial filters used by the model. For denser graphs, it is possible to cover the entire graph using just a few graph convolutional layers (so that each node has information from most if not all of the other nodes). In general, we find that if the average degree is high and the diameter is small, the nodes are more connected and, therefore, fewer convolutional layers are needed.

Beyond the convolutional layer, there are design choices for pooling and aggregation layers. If we add pooling layers and cluster the nodes before the aggregation layer, as in DiffPool \cite{ying2018} and Top-k Pool \cite{gao2019}, we can capture hierarchical structure (see Section~\ref{subsec:graphpoolinglayer-1}). Some other approaches include embedding graphs as a feature (e.g., family of graph spectral distances (F$_{\text{GSD}}$) \cite{verma2017}), and using multiple statistics or attention-based capsules to capture more global information (see Section~\ref{subsec:graphaggregationlayer}). Results are shown in Section~\ref{sec:res} Fig.~\ref{fig:pooling_results} and Fig.~\ref{fig:aggregation_results}.

\section{Results}
\label{sec:res}

\subsection{Graph Convolution}
\label{subsec:graph_convolution_results}

\begin{figure}[ht]
    \centering
    \includegraphics[width=0.4\textwidth]{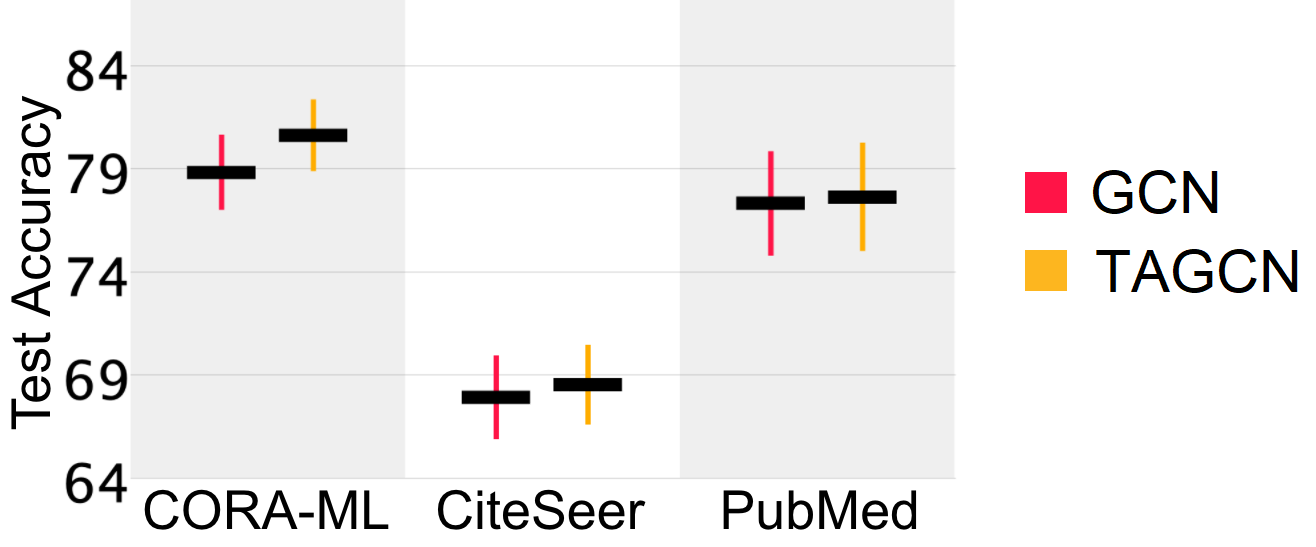}
    \caption{Comparison of graph CNN variants for node classification on benchmark datasets (see \cite{node_dataset} for dataset descriptions and \cite{kipf17} for experimental setup). The horizontal black line represents the mean accuracy and the vertical line represents the standard deviation.}
    \label{fig:node_classification_results}

\end{figure}

We compare GCNs \cite{kipf17} and TAGCNs \cite{du2017} for node and graph classification on popular benchmark datasets in Fig.~\ref{fig:node_classification_results} and Fig.~\ref{fig:pooling_results} (``No pool''), respectively. See \cite{node_dataset,graphstat} for descriptions of datasets. For TAGCN, we consider graph polynomial filters up to degree 3. We perform all experiments using the PyTorch Geometric Library \cite{fey2019}.

In general, TAGCN performs better than GCN for both node and graph classification in terms of mean classification accuracy. However, the TAGCN filtering operation has more complexity, overfitting more easily and may lead to higher variance. For node classification on citation networks, GCN and TAGCN empirically perform poorly with more than 2 layers, as they tend to oversmooth. For graph classification, GCN and TAGCN require more layers to achieve high performance on sparser graphs, as measured by average degree or graph diameter. For a more in-depth discussion, see \cite{du2017}.

\subsection{Graph Pooling}
\label{subsec:graph_pool_results}
\begin{figure}
    \centering
    \includegraphics[width=\linewidth]{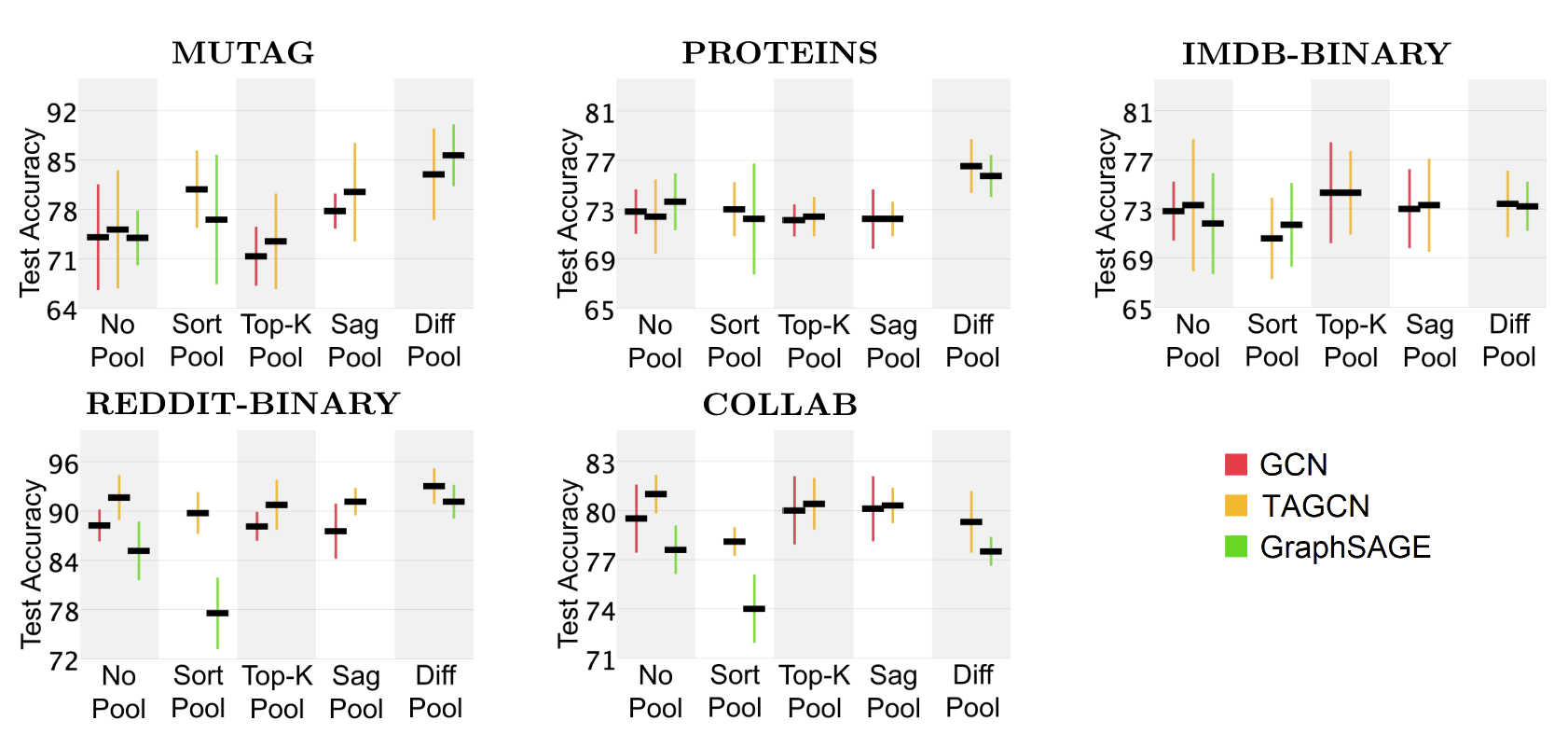}
     \caption{Comparison of pooling methods for graph classification on benchmark datasets (see \cite{graphstat} for dataset descriptions). The horizontal black line represents the mean accuracy and the vertical line represents the standard deviation.}
      \label{fig:pooling_results}

\end{figure}

We compare different pooling algorithms for graph classification used with graph CNNs on popular benchmark datasets (see \cite{graphstat} for descriptions of datasets) in Fig.~\ref{fig:pooling_results} \footnote{We include GraphSAGE in some of our analysis as it is used in some of the original papers.}. In general, pooling helps if the graphs are sparse (MUTAG, PROTEINS, IMDB-B, and REDDIT-B). We conjecture that DiffPool generally performs better on the benchmark datasets because it learns a new graph structure and signal (instead of selecting from existing nodes and edges). SagPool and SortPool perform better for MUTAG and PROTEINS, but they are similar or worse for IMDB-B and REDDIT-B. TAGCN tends to perform better than the other architectures.  For a more in-depth discussion, see \cite{cheung2019}.

\setlength{\belowcaptionskip}{0pt}
\setlength{\belowcaptionskip}{-10pt}
\subsection{Graph Aggregation}

\begin{figure}
     \centering
    \includegraphics[width=\linewidth]{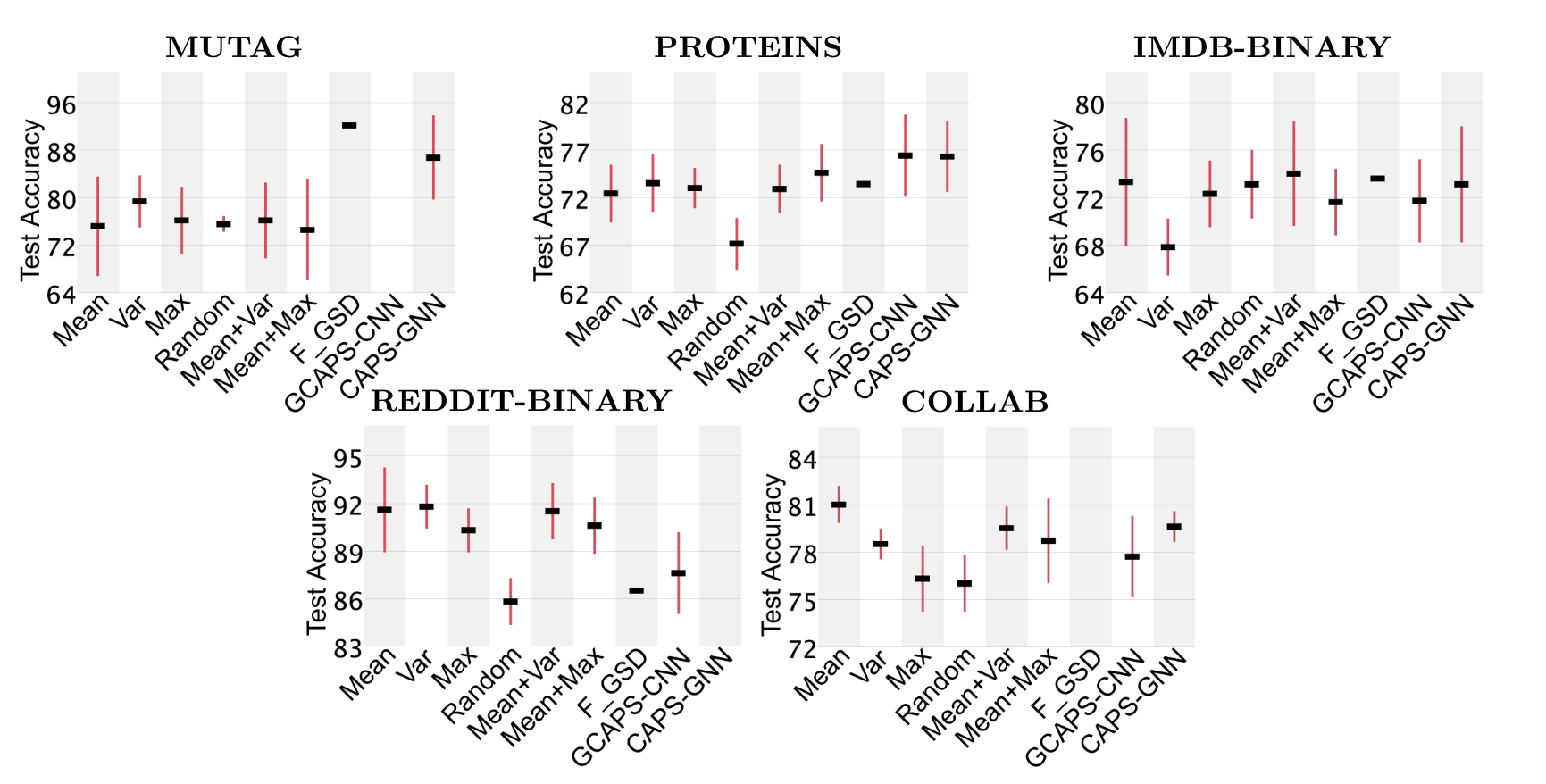}
        \caption{Comparison of aggregation methods for graph classification on benchmark datasets (see \cite{graphstat} for dataset descriptions). The horizontal black line represents the mean accuracy and the vertical red line represents the standard deviation.}
        \label{fig:aggregation_results}
\end{figure}

We compare different graph aggregation methods for graph classification used with graph CNNs on popular benchmark datasets (see \cite{graphstat} for descriptions of datasets) in Fig.~\ref{fig:aggregation_results} \footnotemark \footnotetext{The results for $\textrm{F}_{\textrm{GSD}}$, GCAPS-CNN and CapsGNN are taken from \cite{verma2017}, \cite{verma2018}, and \cite{xinyi2019}, respectively, and blank if no results are provided for given datasets.}. In general, for graph CNNs, there is no single best aggregation method across the considered datasets. However, certain aggregation methods may be complementary (e.g., combining mean and variance leads to an increase in accuracy), but some combinations lead to overfitting (e.g., combining mean and max actually reduces the test accuracy for the IMDB-B dataset). Capsule-based techniques improve the performance for some datasets (e.g.  GCAPS-CNN and CapsGNN improve performance for the MUTAG and PROTEINS datasets). Some datasets are classified well using just the graph structure but not the graph signal (e.g., MUTAG, PROTEINS, IMDB-B show good accuracy with $\textrm{F}_{\textrm{GSD}}$, which does not consider graph signals).

\section{Conclusion}
\label{sec:Conc}
Inference tasks on graph-structured data demand learning architectures that can adequately model non-Euclidean structure, and GSP---e.g., with graph filters of different degrees---provides a principled methodology for the design and analysis of those architectures. As the field continues to develop, new connections will emerge and the formalism of GSP should continue to play a significant role in deep learning on graphs.

The identity matrix case shown in Fig.~\ref{fig:graphEffect} is analogous to ignoring the graph structure and applying a multilayer perceptron (MLP) to the problem. When comparing deep learning architectures, it is typical to compare them in terms of their representational power, but this is insufficient for understanding potential performance. CNNs, for example, are not as general as MLPs, but in practice are far more effective in domains where shift invariance is important. Shift invariance is invaluable for vision-based pattern recognition, but, like Tolstoy's unhappy families\footnote{``Happy families are all alike; every unhappy family is unhappy in its own way,'' from \textit{Anna Karenina} (1877) by Leo Tolstoy.}, no graph problems are alike: there is no guarantee that the properties that make a particular deep learning architecture effective on one class of graph-structured problems apply to another. The data and structure of the problem domain are as important as the learning architecture itself, and this importance only grows in the irregular world of geometric deep learning.

We identify some open research questions on the application of GSP to  deep learning:
\begin{itemize}
    \item \textbf{Different kernels for graph convolution:} GCN and TAGCN define the convolutional layer in terms of the adjacency matrix. \textit{Can we use other kernels?} One candidate is spectral domain convolution, introduced in \cite{shisample}, a polynomial of the spectral domain shift $\textbf{M} = \textrm{GFT} \Lambda^* \textrm{GFT}^{-1}$ where $\Lambda^*$ is the complex conjugate of the matrix of eigenvalues in \eqref{eq:A_eigengsp}.

    \item \textbf{When graph CNNs fail:}
 Popular graph CNNs can incorrectly produce the same output for non-isomorphic graphs \cite{XuHLJ19}. GCN and TAGCN work well in practice, but fail on certain edge cases. \textit{What approaches can we take to prevent graph CNN failure in these cases? How can we increase the expressive power of graph CNNs?}
 
    \item \textbf{Interpretability of graph pooling methods:} \label{sec:graphsample}
    Reference \cite{shisample} presents sampling methods in both the vertex domain and the spectral domain. \textit{How can we relate aforementioned pooling algorithms (e.g., DiffPool, SagPool) to GSP sampling theory? How can we use GSP sampling theory to develop novel pooling methods?}
    
    \item \textbf{Other Problems:} GSP can also be used to tackle other research directions for graph CNNs, including model depth, oversmoothing \cite{oversmooth}, scalability, heterogenity, dynamicity, and interpretability (see \cite{review1, review2} for more details).
\end{itemize}
{
\pagebreak

\paragraph{Authors:}  \mbox{} \\
\textbf{Mark Cheung} (\href{mailto:markcheu@andrew.cmu.edu}{markcheu@andrew.cmu.edu}) received his B.S. degrees in Electrical Engineering and Computer Engineering in 2014 from University of Virginia. He is currently a Ph.D. candidate in the Electrical and Computer Engineering Department at Carnegie Mellon University. His research interests include geometric deep learning and neuroscience.

\noindent \textbf{John Shi} (\href{mailto:jshi3@andrew.cmu.edu}{jshi3@andrew.cmu.edu}) received his B.S. degrees in Computer Engineering and Applied Mathematics in 2017 from the University of Maryland, College Park. He is currently a Ph.D. candidate in the Electrical and Computer Engineering Department at Carnegie Mellon University. His research interests include graph signal processing theory and applications.

\noindent \textbf{Oren Wright} (\href{mailto:owright@sei.cmu.edu}{owright@sei.cmu.edu}) received his B.S. degree in 2010 and his M.S. degree in 2011 in Electrical and Computer Engineering from Carnegie Mellon University. In 2016, he joined the Carnegie Mellon University Software Engineering Institute, where he has since led research in machine learning and its application to critical defense problems.

\noindent \textbf{Lavender Yao Jiang} (\href{mailto:lavenderjiang@cmu.edu}{lavenderjiang@cmu.edu}) is an undergraduate student in the Electrical and Computer Engineering Department and the Mathematical Sciences Department at Carnegie Mellon University. Her research interests include evaluating the quality of graphs and biomedical applications of graph signal processing. 

\noindent \textbf{Xujin Liu} (\href{mailto:xujinl@andrew.cmu.edu}{xujinl@andrew.cmu.edu}) is an undergraduate student in the Electrical and Computer Engineering Department at Carnegie Mellon University. His research interests include graph theory, graph neural networks, and their application to biomedical fields.

\noindent \textbf{Jos\'e M.F. Moura} (\href{mailto:moura@andrew.cmu.edu}{moura@andrew.cmu.edu}) is the Philip L. and Marsha Dowd University Professor at Carnegie Mellon University. The technology of two of his patents (co-inventor Alek Kavcic) are in over four billion disk drives in 60\% of all computers sold worldwide in the last 15 years. He is Fellow of IEEE, AAAS, and U.S. National Academy of Inventors, corresponding member of Portugal Academy of Sciences, and member of U.S. National Academy of Engineers. He holds a Doctor Honoris Causa from University of Strathclyde, UK, and received the Grand Cross of the Order of Prince Henry from the President of the Republic of Portugal. He was 2019 IEEE President and CEO. }

\pagebreak
\singlespacing
\bibliographystyle{IEEEtran}
\bibliography{IEEEabrv,references}
\end{document}